\documentclass[twocolumn,pra,superscriptaddress,floatfix,showpacs]{revtex4-1}

\usepackage{hyperref}
\usepackage{graphicx}
\usepackage{amsmath}
\usepackage{color}
\usepackage{units}
\usepackage[latin1]{inputenc}

\begin{document}

\title{Thermal Noise Reduction and Absorption Optimisation via Multi-Material Coatings}

\author{Jessica Steinlechner}
\affiliation{SUPA, School of Physics and Astronomy, University of Glasgow, Glasgow,
G12 8QQ, Scotland}
\affiliation{Institut f\"ur Gravitationsphysik, Leibniz Universit\"at Hannover and Max-Planck-Institut f\"ur Gravitationsphysik (Albert-Einstein-Institut), Callinstr. 38, 30167 Hannover, Germany}
\author{Iain W Martin}
\email{iain.martin@glasgow.ac.uk}
\affiliation{SUPA, School of Physics and Astronomy, University of Glasgow, Glasgow,
G12 8QQ, Scotland}
\author{Jim Hough}
\affiliation{SUPA, School of Physics and Astronomy, University of Glasgow, Glasgow,
G12 8QQ, Scotland}
\author{Christoph Kr\"uger}
\affiliation{Institut f\"ur Gravitationsphysik, Leibniz Universit\"at Hannover and Max-Planck-Institut f\"ur Gravitationsphysik (Albert-Einstein-Institut), Callinstr. 38, 30167 Hannover, Germany}
\author{Sheila Rowan}
\affiliation{SUPA, School of Physics and Astronomy, University of Glasgow, Glasgow,
G12 8QQ, Scotland}
\author{Roman Schnabel}
\affiliation{Institut f\"ur Gravitationsphysik, Leibniz Universit\"at Hannover and Max-Planck-Institut f\"ur Gravitationsphysik (Albert-Einstein-Institut), Callinstr. 38, 30167 Hannover, Germany}
\affiliation{Institut f\"ur Laserphysik and Zentrum f\"ur Optische Quantentechnologien, Universit\"at Hamburg, Luruper Chaussee 149, 22761 Hamburg, Germany}

\begin{abstract}
Future gravitational wave detectors (GWDs) such as Advanced LIGO upgrades and the Einstein Telescope are planned to operate at cryogenic temperatures using crystalline silicon (cSi) test-mass mirrors at an operation wavelength of 1550\,nm. The reduction in temperature in principle provides a direct reduction in coating thermal noise, but the presently used coating stacks which are composed of silica (SiO$_2$) and tantala (Ta$_2$O$_5$) show cryogenic loss peaks which results in less thermal noise improvement than might be expected. Due to low mechanical loss at low temperature amorphous silicon (aSi) is a very promising candidate material for dielectric mirror coatings and could replace Ta$_2$O$_5$. Unfortunately, such a aSi/SiO$_2$ coating is not suitable for use in GWDs due to high optical absorption in aSi coatings. We explore the use of a three material based coating stack. In this multi-material design the low absorbing Ta$_2$O$_5$ in the outermost coating layers significantly reduces the incident light power, while aSi is used only in the lower bilayers to maintain low optical absorption. Such a coating design would enable a reduction of Brownian thermal noise by \unit[25]{\%}. We show experimentally that an optical absorption of only $\unit[(5.3 \pm 0.4)]{ppm}$ at 1550\,nm should be achievable.
\end{abstract}


\maketitle

\section{Introduction}

Future gravitational wave detectors (GWDs) such as Advanced LIGO upgrades and the low frequency (LF) detector within the Einstein Telescope~\cite{ETdesign, Hild2010} (ET) are planned to operate at cryogenic temperature to reduce thermal noise. Operation at low temperatures requires a replacement substrate material for the presently used fused silica test-mass mirrors. Showing low mechanical loss at low temperatures~\cite{Nawrodt2008}, crystalline silicon (cSi) is planned to be used at an operation wavelength of \unit[1550]{nm} to keep optical absorption low~\cite{Keevers1995}. \unit[120]{K} and \unit[20]{K} are interesting operation temperatures for cSi with low thermo-elastic noise due to zero crossings of the thermal expansion coefficient $a_{\rm th}$. For ET an even lower operation temperature of \unit[10]{K} is planned~\cite{ETdesign}.

Coating thermal noise power spectral density is also proportional to the temperature of the coating, and so a reduction in temperature should in principle provide a direct reduction in thermal noise. However, the mechanical loss of many materials is temperature dependent, and cryogenic loss peaks have been identified in a doped Ta$_2$O$_5$/SiO$_2$ coating stack~\cite{Granata2013} and in single layers of SiO$_2$~\cite{Martin2014} and Ta$_2$O$_5$~\cite{Martin2009}.
As a result of this increase in loss at low temperature, there is less thermal noise improvement than might be expected from operation at cryogenic temperature. Therefore, low mechanical loss at low temperature makes amorphous silicon (aSi) a very promising candidate material for dielectric mirror coatings and could replace Ta$_2$O$_5$ in the presently used coatings as a first step in improving the overall coating loss.

A standard highly reflective (HR) quarter-wavelength coating is composed of a stack of alternating materials of differing refractive indices, where each layer has an optical thickness $\delta = n \times t =\lambda/4$ (where t is the geometric thickness) for the wavelength of interest. The reflectivity depends on the number of bilayers in the coating, and on the difference in refractive index between the two materials. Thus, for a desired reflectivity, the total thickness of coating required depends on the difference in refractive index between the materials used.

Presently, stacks of alternating layers of SiO$_2$ and Ta$_2$O$_5$ are used as highly reflective coatings for GWD test-mass mirrors. aSi has a significantly higher refractive index than Ta$_2$O$_5$ ($n_{\rm Si}=3.5$ and $n_{\rm Ta_2O_5}=2.2$ at \unit[1550]{nm}). Thus in comparison to a Ta$_2$O$_5$/SiO$_2$ coating, a coating formed from aSi/SiO$_2$ will have thinner high-index layers, and will achieve the same reflectivity with fewer bilayers. While a Ta$_2$O$_5$/SiO$_2$ coating stack requires 18 bilayers (plus a $\lambda/2$ SiO$_2$ protection layer and a $\lambda/4$ Ta$_2$O$_5$ transitional layer) to achieve a high reflectivity with a transmission of $T \approx \unit[0.5]{ppm}$, a aSi/SiO$_2$ stack needs only 8 bilayers to achieve equivalent reflectivity.
This significantly reduces the total thickness of the coating stack from $\unit[8.7]{\mu m}$ to $\unit[3.7]{\mu m}$ and provides a direct reduction in coating thermal noise. Furthermore, the cryogenic mechanical loss of aSi is at least a factor of 10 lower than that of Ta$_2$O$_5$, providing another direct reduction in coating thermal noise.

However preliminary studies of the optical absorption of a highly reflective Si/SiO2 coating stack at \unit[1550]{nm} have shown an absorption level of approximately \unit[1000]{ppm}~\cite{Silicon_Coatings} which is substantially above the coating absorption requirement of future GWDs (which is in the order of \unit[1]{ppm}). Thus optical absorption in aSi coatings would need be to significantly reduced for such an two-material arrangement containing aSi to be implemented

In this paper we explore the use of multi-material coatings and use measurements of aSi, Ta$_2$O$_5$ and SiO$_2$ to demonstrate the optical and thermal noise performance of a three material coating. A more general theory of multi-material coatings, derived independently and in parallel to the work presented here, is given by Yam, Gras and Evans~\cite{Evans2014}. Here we investigate a three material design in which low absorption Ta$_2$O$_5$ (combined with SiO$_2$) is used as the high-index material in the outermost 7 bilayers (plus a $\lambda/2$ SiO$_2$ protection layer and a $\lambda/4$ Ta$_2$O$_5$ transitional layer), which reflect more than $\unit[99.65]{\%}$ of the incident light power, while aSi is used as the high-index material in the 5 lower bilayers, which can tolerate the higher optical absorption due to the relatively low light power present.
This design should in principle allow the higher refractive index and lower mechanical loss of aSi to be exploited, without significantly reducing the optical performance of the coating stack.

\section{Model of the multi-material coating stack}

\begin{figure}
  \centering
  \includegraphics[width=7cm]
    {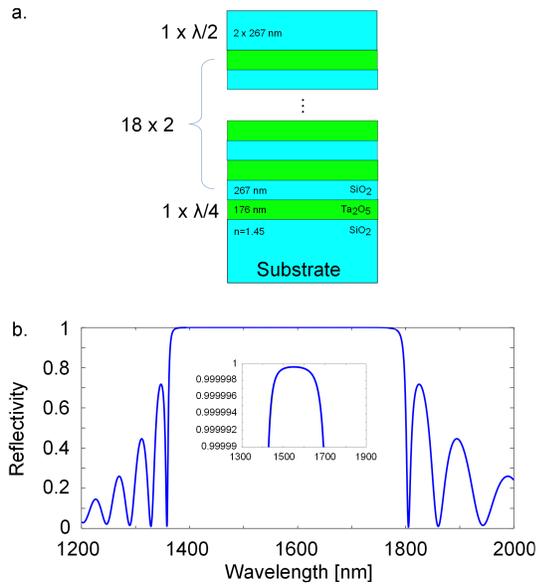}
  \caption{a. Schematic of a coating stack which consists of 18 $\lambda/4$ bilayers of alternating SiO$_2$ and Ta$_2$O$_5$ plus a $\lambda/2$ SiO$_2$ protection layer on the surface and a $\lambda/4$ Ta$_2$O$_5$ transition layer. The quarter layer thicknesses are $t_{\rm SiO_2}=\unit[267]{nm}$ and $t_{\rm Ta_2O_5}=\unit[176]{nm}$ resulting a total thickness of $\unit[8.684]{\mu m}$. b. The transmission of this stack versus wavelength. At \unit[1550]{nm} the transmission is $T \approx \unit[0.5]{ppm}$.}
  \label{fig:stack_sio2_ta2o5}
\end{figure}

For the arm cavity end test mass (ETM) in cryogenic ET it is planned to use coating stacks of 18 $\lambda/4$ bilayers of alternating SiO$_2$ and Ta$_2$O$_5$~\cite{ETdesign}. The refractive indices at \unit[1550]{nm} are $n_{\rm SiO_2}=1.45$ and $n_{\rm Ta_2O_5}=2.2$. At each layer boundary the change of refractive index causes a reflection following Fresnel's equations~\cite{Fresnel}. Using ray-transfer matrix formalism for calculating a series of reflecting boundaries, the actual reflectivity within each bilayer of the coating stack was calculated (for details see~\cite{Steini2013}). Additionally to 18 $\lambda/4$ bilayers, our model stack has a $\lambda/2$ protection layer of SiO$_2$ on the outside surface and a $\lambda/4$ Ta$_2$O$_5$ transitional layer between substrate and the first SiO$_2$/Ta$_2$O$_5$ bilayer starting with SiO$_2$. The total reflectivity of the stack is $R=\unit[99.99995]{\%} \, (T= 1-R \approx \unit[0.5]{ppm})$. The total thickness of the stack is $\unit[8.684]{\mu m}$ of which $\unit[5.340]{\mu m}$ consist of SiO$_2$ and $\unit[3.344]{\mu m}$ of Ta$_2$O$_5$. A schematic of this model stack is shown in Fig.~\ref{fig:stack_sio2_ta2o5}(a), while Fig.~\ref{fig:stack_sio2_ta2o5}(b), shows the reflectivity of the coating stack versus wavelength.

\begin{figure}
  \centering
  \includegraphics[width=7cm]
    {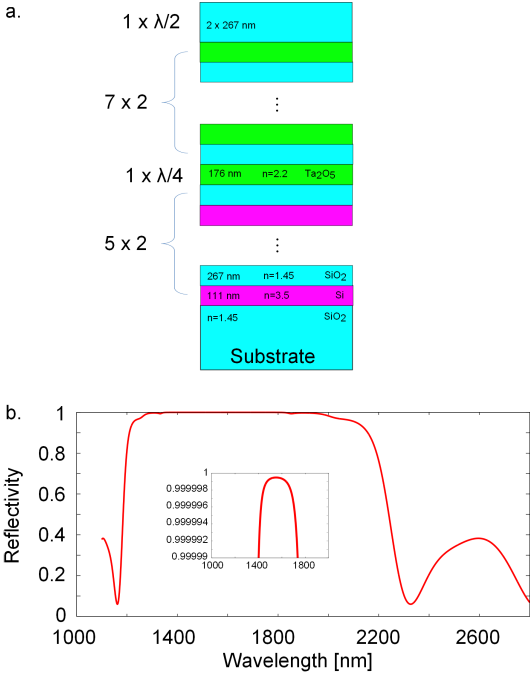}
  \caption{a. Schematic of the multi-material coating stack. The stack consists of 7 $\lambda/4$ bilayers of alternating SiO$_2$ and Ta$_2$O$_5$ plus a $\lambda/2$ SiO$_2$ protection layer on the surface and a $\lambda/4$ Ta$_2$O$_5$ transition layer. Below this stack another 5 $\lambda/4$ bilayers of alternating aSi and SiO$_2$ follow. The quarter layer thicknesses are $t_{\rm SiO_2}=\unit[267]{nm}$, $t_{\rm Ta_2O_5}=\unit[176]{nm}$ and $t_{\rm Si}=\unit[111]{nm}$ resulting in a total thickness of $\unit[5.701]{\mu m}$ for the multi-material coating stack. b. The transmission of the multi-material stack versus wavelength. Equivalent to the SiO$_2$/Ta$_2$O$_5$ stack the transmission at \unit[1550]{nm} is $T \approx \unit[0.5]{ppm}$.}
  \label{fig:multi_material}
\end{figure}

In a next step we explore a coating stack which starts with 7 bilayers of SiO$_2$/Ta$_2$O$_5$ (plus a $\lambda/2$ protection layer of SiO$_2$ and a Ta$_2$O$_5$ transitional layer) to reflect most of the laser light. Only \unit[0.35]{\%} of laser light is transmitted by this stack. Additionally, 5 bilayers of aSi/SiO$_2$ are used to reduce the total transmission to $T \approx \unit[0.5]{ppm}$, see Fig.~\ref{fig:multi_material}. This reduces the number of single layers from 38 to 26 while the total thickness of the stack is reduces from $\unit[8.684]{\mu m}$ to $\unit[5.701]{\mu m}$ (by \unit[34]{\%}). In Tab.~\ref{tab:thickness} the total thickness of the coating stacks and of each material are listed.

\section{Thermal noise of the coating stack}

The power spectral density of the coating thermal noise is proportional to both the mechanical loss and the total thickness of the coating stack as shown in Eq.~\ref{eqn:Harry_big}~\cite{Harry2002}

\begin{eqnarray}
 S_x(f)& = & \frac{2k_BT}{\pi^{3/2}f}\frac{1-\sigma^2}{w_0Y}\Big\{\phi_{\textrm{substrate}}\nonumber\\
 &&+\frac{1}{\sqrt{\pi}} \frac{d}{w_0}\frac{1}{YY^\prime(1-\sigma^\prime2)(1-\sigma^2)}\nonumber\\
 &&\times {} [Y^{\prime2}(1+\sigma)^2(1-2\sigma)^2\phi_\parallel\nonumber\\
 &&+{}YY^\prime\sigma^\prime(1+\sigma)(1+\sigma^\prime)(1-2\sigma)(\phi_\parallel-\phi_\perp)\nonumber\\
 &&+{}Y^2(1+\sigma^\prime)^2(1-2\sigma^\prime)^2\phi_\perp]\Big\},
\label{eqn:Harry_big}
\end{eqnarray}

where $f$ is the frequency in Hz, $T$ is the temperature in Kelvin, $Y$ and $\sigma$ are the Young's modulus and Poisson's ratio of the substrate, $Y'$ and $\sigma^\prime$ are the Young's modulus and Poisson's ratio of coating, $\phi_\parallel$ and $\phi_\perp$ are the mechanical loss values for the coating for strains parallel and perpendicular to the coating surface, $d$ is the coating thickness and $w_0$ is the laser beam waist.

To estimate the thermal noise associated with this coating stack, it is necessary to first estimate the mechanical loss of the stack. This can be calculated as a weighted (by thickness and by stiffness / Young's modulus) average of the loss of the individual coating materials as shown in Eqs~\ref{eq:phi} and~\ref{eq:y}~\cite{Jones1975}. To calculate the loss of the 38 layer SiO$_2$/Ta$_2$O$_5$ coating stack and of the 26 layers the SiO$_2$/Ta$_2$O$_5$/aSi multi-material stack the loss values given in Tab.~\ref{tab:loss} were used.
 
\begin{equation}
	Y_{\rm coating} = \frac{Y_{\rm SiO_2}t_{\rm SiO_2} + Y_{\rm Ta_2O_5}t_{\rm Ta_2O_5} (+Y_{\rm aSi}t_{\rm aSi})}{t_{\rm SiO_2} + t_{\rm Ta_2O_5} (+t_{\rm aSi})}
	\label{eq:y}
\end{equation}

\begin{eqnarray}
	\phi_{\rm coating} & = & \frac{Y_{\rm SiO_2}t_{\rm SiO_2}\phi_{\rm SiO_2} + Y_{\rm Ta_2O_5}t_{\rm Ta_2O_5}\phi_{\rm Ta_2O_5}}{Y_{\rm coating}t_{\rm coating}}\nonumber\\
	&& \left( + \frac{Y_{\rm Si}t_{\rm aSi}\phi_{\rm aSi}}{Y_{\rm coating}t_{\rm coating}} \right)
\label{eq:phi}
\end{eqnarray}

\begin{figure}
  \centering
  \includegraphics[width=7cm]{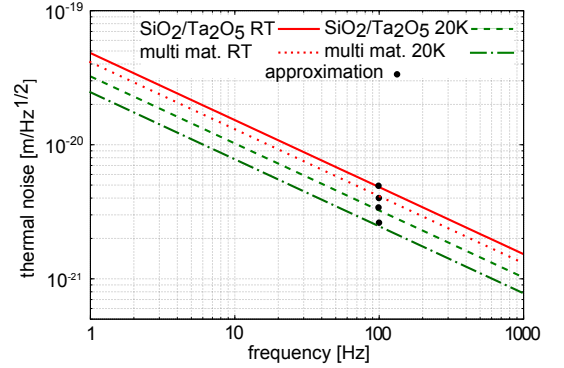}
  \caption{Brownian thermal noise for the SiO$_2$/Ta$_2$O$_5$ coating at room temperature (red line) and at \unit[20]{K} (dashed green line) and for the multi-material coating at room temperature (dotted red line) and at \unit[20]{K} (dash-dotted green line) calculated from the model used in~\cite{Evans2014}. The black dots show the corresponding results calculated with Eq.~\ref{eqn:Harry_big}.}
  \label{fig:Brownian_thermal_noise}
\end{figure}

Equation~\ref{eq:phi} gives the total loss $\phi_{\rm coating}$ which is composed of the loss of different materials, while Eq.~\ref{eq:y} allows the Young's modulus of the components to be calculated~\cite{Crooks2003}. The thickness per material used for the calculations can be found in Tab.~\ref{tab:thickness}. The results for the loss are summarized in Tab.~\ref{tab:loss}. While at room temperature (RT) the loss for the multi-material stack is slightly higher than for the SiO$_2$/Ta$_2$O$_5$ stack, at lower temperatures due to the decreasing aSi loss the loss of the multi-material stack becomes lower. This results in a reduction of Brownian thermal noise which was calculated using Eq.~\ref{eqn:Harry_big}. The results are also given in Tab.~\ref{tab:loss} and represented by the black dots in Fig~\ref{fig:Brownian_thermal_noise}. This approximation is in good agreement with the more precise model used in~\cite{Evans2014} with discrepancies of less than \unit[5]{\%}. Figure~\ref{fig:Brownian_thermal_noise} also shows the Brownian thermal noise for the SiO$_2$/Ta$_2$O$_5$ (lines) coating and the multi-material coating (crosses and pluses) at RT (red, upper two curves) and at \unit[20]{K} (green, lower two curves) calculated from the model used in~\cite{Evans2014}.

\begin{table}
\centering
  \caption{Thickness t for the SiO$_2$/Ta$_2$O$_5$ coating stack, for the multi-material stack (SiO$_2$/Ta$_2$O$_5$/aSi) and for the individual materials, and Young's modulus Y of the coatings and individual materials.}
 \label{tab:thickness}
 \begin{tabular}{llllll}
        \hline
																		&\multicolumn{2}{l}{SiO$_2$/Ta$_2$O$_5$}	&\multicolumn{3}{l}{multi mat.}\\
				t$_{\rm total}$ [$\mu$m]		&\multicolumn{2}{l}{8.684}								&\multicolumn{3}{l}{5.701}\\
				Y$_{\rm coating}$ [GPa]			&\multicolumn{2}{l}{98}										&\multicolumn{3}{l}{96}\\
				\hline
																		&SiO$_2$ 			&Ta$_2$O$_5$								&SiO$_2$ 			&Ta$_2$O$_5$		&aSi\\
				\hline
				t								[$\mu$m]	  &5.34					&3.344											&3.738				&1.408					&0.555\\			
				Y								[GPa]				&72						&147												&							&  							&140\\										
				\hline				
 \end{tabular}
 \end{table}

\begin{table*}
\centering
  \caption{Loss $\phi$ for SiO$_2$, Ta$_2$O$_5$ and aSi, and resulting loss for a SiO$_2$/Ta$_2$O$_5$ stack and the multi-material stack each at room temperature (RT), \unit[120]{K}, \unit[20]{K} and \unit[10]{K}. The resulting Brownian thermal noise for the two stacks at each temperature was calculated using Eq.~\ref{eqn:Harry_big}.}
 \label{tab:loss}
 \begin{tabular}{llllllll}
        \hline
				&\multicolumn{5}{l}{loss $\phi$ $\times 10^{-4}$}																						&\multicolumn{2}{l}{Brownian th. noise (\unit[100]{Hz}) $\times 10^{-21}$}\\
				\hline							
																&SiO$_2$   								&aSi 										&Ta$_2$O$_5$					&SiO$_2$/Ta$_2$O$_5$	&multi mat.	&SiO$_2$/Ta$_2$O$_5$				&multi mat.\\
 				\hline
 				\unit[290]{K}						& 0.4~\cite{Flaminio2010}	&4.0~\cite{Murray2015}	&2.3~\cite{Flaminio2010}&1.26											&1.62	& 4.9		&4.3\\
				\unit[120]{K}						& 1.7~\cite{Martin2014}		&0.5~\cite{Murray2015}	&3.3~\cite{Martin2010} 	&2.58												&2.1	& 4.5		&3.5\\	
        \unit[20]{K}						& 7.8~\cite{Martin2014}		&0.4~\cite{Murray2015}	&8.6~\cite{Martin2010}	&8.24												&6.98	& 3.4		&2.6\\
        \unit[10]{K} 						& 7~\cite{Martin2014}			&0.3~\cite{Murray2015}	&6~\cite{Martin2010}		&6.46												&5.64	& 2.2		&1.7\\
 				\hline				
 \end{tabular}
 \end{table*}

\section{Optical absorption of the Coating Stack}

To estimate the optical absorption of this multi-material stack, we consider two experiments in which we measured the absorption of a SiO$_2$/Ta$_2$O$_5$ coating stack and a aSi/SiO$_2$ coating stack separately using \textit{Photothermal Self-Phase Modulation} (PSM)~\cite{SHG}. In Subsec.~\ref{subsec:tantala} we present measurements of the optical absorption of a SiO$_2$/Ta$_2$O$_5$ coating stack at \unit[1550]{nm} for the first time, while in Subsec.~\ref{subsec:silicon} a new aspect of the absorption results of a aSi/SiO$_2$ coating stack are discussed which were published in~\cite{Silicon_Coatings}.

\subsection{Silica/Tantala}
\label{subsec:tantala}

For the absorption measurement on SiO$_2$/Ta$_2$O$_5$ at \unit[1550]{nm} a Fabry-Pérot cavity was used. The two cavity mirror substrates consisted of Corning 7980 glass~\cite{corning}. The coatings were manufactured at \textit{Advanced Thin Films} (ATF)~\cite{ATF} and were optimized for a finesse of 10 000 at an angle of incidence (AOI) of $0^\circ$ at \unit[1550]{nm}. A finesse of 10 000 theoretically is reached by two identical lossless mirrors with intensity reflectivity of R$_1$=R$_2$=\unit[99.969]{\%} (1-R=\unit[310]{ppm}). The transmission measured by the manufacturer on a mirror from this coating run was T=\unit[320]{ppm}.
%
%
\subsubsection{Experimental Setup}

For the Fabry-Pérot cavity setup, two identical curved mirrors M$_1$ and M$_2$ were clamped facing each other, separated only by a \unit[0.75]{mm} viton seal. A PZT, driven by a function generator (FG), presses M$_2$ into the viton seal and therefore changes the cavity length. Both mirrors have a concave radius of curvature (ROC) of \unit[0.5]{m} resulting in a cavity waist of $w_0 = \unit[82]{\mu m}$. The short cavity length results in a large free spectral range (FSR) of \unit[200]{GHz}. The material properties and cavity geometry parameters of this setup are summarized in Tab.~\ref{tab:Coating_parameters_1550nm}.

\begin{table*}
\centering
  \caption{Material and geometric parameters of the Corning 7980 mirror substrates and the cavity, as used for the simulations. Wavelength and temperature dependent parameters are given at $\unit[1550]{nm}$ and room temperature.}
 \label{tab:Coating_parameters_1550nm}
 \begin{tabular}{lllll}
        \hline
        Material parameters			 					& 																	&Ref.										&Cavity geometry parameters		&\\
 				\hline
 				index of refraction $n$						& $1.44$ 														&\cite{leviton06}				&input power $P$				&\unit[2, 130, 260]{mW}\\
        thermo refr. coeff. d$n$/d$T$			& $\unit[9.6\cdot 10^{-6}]{/K}$			&\cite{corning}					&length $L$							&\unit[0.75]{mm}\\
        specific heat $c$ 								& $\unit[770]{J/(kg\,K)}$ 					&\cite{val}							&beam waist $w_0$				&\unit[82]{$\mu$m}\\
        density $\rho$ 										& $\unit[2201]{kg/m^3}$ 						&\cite{corning}					&mirror thickness $D$		&\unit[6.35]{mm}\\
        thermal expansion $a_{\rm th}$ 		& $\unit[0.52\cdot 10^{-6}/]{K}$ 		&\cite{corning}					&angle of incid. (AOI)	&$0^\circ$\\
        thermal conductivity $k_{\rm th}$ & $\unit[1.3]{W/(m\,K)}$ 						&\cite{corning}					&												&\\
 				\hline				
 \end{tabular}
 \end{table*}

To linearize the mirror motion, the cavity length was modulated only in a small range around the resonance of approximately \unit[0.3]{\%} of a FSR: this range varied slightly with the modulation frequency (the modulation voltage was constant for all measurements). The actual mirror motion for each modulation frequency and expansion and contraction of the PZT in each case was calibrated using side bands imprinted on the laser signal via an electro optical modulator (EOM) before entering the cavity, for details see~\cite{Thesis}. The resonance peaks for the absorption measurement were detected with a photo detector (PD) in reflection of input mirror $M_1$. The reflected beam was separated from the incoming beam at a polarizing beam splitter cube (PBS) using a $\lambda/2$ waveplate and a Faraday Rotator.
%
%
\subsubsection{Results}

\begin{figure}
  \centering
  \includegraphics[width=6cm]{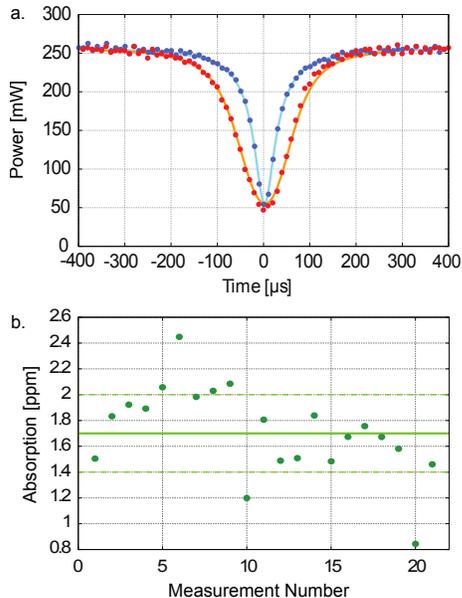}
  \caption{Measurement results for tantala HR-coatings at \unit[1550]{nm}: In a. an example of deformed resonance peaks is shown from which the coating absorption was calculated. For an external lengthening of the cavity the resonance peak is broadened (simulation: orange line, measurement: red dots) while for an external shortening of the cavity the peak is narrowed (simulation: light blue line, measurement: dark blue dots). b. shows the results for the absorption of the single measurements (dark-green dots). The light-green line and dashed lines mark the mean value and standard deviation of $\alpha=\unit[(1.7 \pm 0.3)]{ppm}$.}
  \label{fig:coatings_1550nm_results}
\end{figure}

Resonance peaks detected at fast scan frequency or low power show no thermal effect (resonance peaks identical for lengthening and shortening of the cavity). Such measurements allow a determination of the reflectivity R$_1$ of the input mirror M$_1$ and the effective reflectivity $\tilde{R}_2$ of $M_2$ which contains all round-trip losses. Assuming R$_1$=R$_2$, the cavity round-trip loss is $\tilde{R}_2-R_1$. A reduction of the scan frequency and a power increase cause a thermal effect (resonance peaks are different for lengthening and shortening of the cavity as shown in Fig.~\ref{fig:coatings_1550nm_results} a.). These measurement additionally allow the determination of the absorption coefficient M$_1$. To measure the absorption, several single measurements of reflected resonance peaks were detected, partly showing a thermal effect, partly showing no thermal effect. The measured peaks were fitted as described in~\cite{Stei12} with the input parameters shown in Tab.~\ref{tab:Coating_parameters_1550nm}.

Altogether 34 single measurements were performed, 21 showing a significant thermal effect. The results for the reflectivities were $1-R_1=\unit[(253.73 \pm 16.4)]{ppm}$ and $1-\tilde{R}_2=\unit[(276.3 \pm 13.3)]{ppm}$ resulting in a cavity finesse of $F=(11853 \pm 704$) which was slightly higher than specified.

The results of the single measurements are shown in Fig.~\ref{fig:coatings_1550nm_results} b., the light green bar marking the mean value of the 21 single results, the dashed light green bars their standard deviation. The resulting absorption for the input mirror coating is $\alpha=\unit[(1.7 \pm 0.3)]{ppm}$ ($\unit[0.3]{ppm} \, \hat{=} \, \unit[17.6]{\%}$ of $\alpha$). This is a promising low result for the optical absorption of SiO$_2$/Ta$_2$O$_5$ at \unit[1550]{nm} and, considering the thicker quarter wave layers due to the longer wavelength, in good agreement with absorption values at \unit[1064]{nm}. (Note that this coating stack was not specified particularly for low absorption. Therefore, further absorption reduction might be possible.)

%
%
\subsubsection{Error Propagation}

For a detailed discussion of the possible error caused by the input parameters see~\cite{Thesis}, where the expected error due to inaccurate material properties is analyzed. Since the material properties are identical or at least very similar to the parameters discussed in~\cite{Thesis}, their influence on the result also is very similar.
The mirror dimensions are also identical to the experiment in~\cite{Thesis}, input power and mode matching always influence the result approximately linear. The only parameters which differ strongly from the other experiment are the cavity length and the beam waist. These two parameters were changed each by $\unit[\pm 10]{\%}$ and the results of one single measurement were recalculated. An error of $\unit[\pm 10]{\%}$ in the cavity length influences the result by approximately $\unit[\pm 10]{\%}$, where a cavity length assumed too long reduces the necessary absorption for the present thermal effect and inverse for a cavity length assumed too short. An error in the cavity length also affects the waist. An error of $\unit[\pm 10]{\%}$ in the length only affects the waist by approximately $\unit[\pm 3]{\%}$, which causes an error of $\unit[\pm 6]{\%}$ in the absorption result. The large error of about \unit[10]{\%} for such a short cavity is realistic. Since cavity length and waist depend on each other as discussed, both errors add. The summed up error is still within the standard deviation for the absorption result.

Therefore for the results obtained from this series of measurements the standard deviation caused by statistical fluctuations of the detected resonance peaks is assumed to dominate the error caused by inaccurate input parameters.
%
%
\subsection{aSilicon/Silica}
\label{subsec:silicon}

Results as well as experimental details for the optical absorption of aSi/SiO$_2$ coatings measured with PSM were published in~\cite{Silicon_Coatings}. The coatings were produced by \textit{Tafelmaier} using \textit{Ion Plating}~\cite{tafelmaier}. In this experiment, the cavity consisted of three mirrors. The input and output mirrors M$_1$ and M$_2$ were identical and reflected the laser beam at an AOI of $42^\circ$, while the third mirror M$_3$ was highly reflective (transmission negligible) at an AOI of $3^\circ$. The optical absorption of the input mirror of the aSi/SiO$_2$ cavity was $\alpha_{\rm M_1,p-pol}=\unit[(1428 \pm 97)]{ppm}$ for p-polarisation and $\alpha_{\rm M_1,s-pol}=\unit[(1035 \pm 42)]{ppm}$ for s-polarisation. The difference in absorption for the two polarisations originates from the difference in reflectivity and therefore different penetration depth of the incident laser beam into the coating stack. Considering a reduction of the laser light due to SiO$_2$/Ta$_2$O$_5$ bilayers to \unit[0.35]{\%}, in s-polarisation \unit[3.6]{ppm} of the input laser light would be absorbed (and \unit[5]{ppm} in p-polarisation) which is in the same order as the absorption of SiO$_2$/Ta$_2$O$_5$ coatings.

\begin{table}
\centering
  \caption{Results for the optical absorption $\alpha$, input mirror reflectivity $R_1$, and effective output reflectivity $\tilde{R}_{2}$ (includes all optical round-trip loss apart from (1-$R_1$)) resulting in an upper limit for the optical absorption $\alpha_{M_3}$=$R_1-\tilde{R}_{2}- 2 \times \alpha_{1}$ of M$_3$.}
 \label{tab:abs}
 \begin{tabular}{lll}
        \hline
				parameter																				&result s-pol									&result p-pol\\
 				\hline
				1- R$_1$ [ppm]																	&529 $\pm$ 20		&4717 $\pm$ 212\\
				$1-\tilde{R}_{2}$ [ppm]													&2902 $\pm$ 103	&7882 $\pm$ 251\\
				$\alpha_{\rm M_1}$ ($=\alpha_{\rm M_2}$) [ppm]	&1035 $\pm$ 42	&1428 $\pm$ 97\\
				$\alpha_{\rm M_3}$ [ppm]												&303 $\pm$ 207	&309 + 657 / -309\\
 				\hline				
 \end{tabular}
 \end{table}

It was observed before that the optical absorption of dielectric coating stacks at a large AOI can be significantly higher than at $0^\circ$. (For a SiO$_2$/Ta$_2$O$_5$ coating stack at an AOI of $42^\circ$, in~\cite{Stei12} a rather high absorption of \unit[23]{ppm} was measured using PSM and confirmed independently with a calorimetric measurement while due to manufacturer approximately \unit[1]{ppm} was expected as demonstrated in~\cite{Lalezari1992}.) Therefore, here we will discuss the optical absorption of mirror M$_3$ which reflects almost at normal incidence from the cavity-round trip loss which is a result provided by PSM in addition to the measurement of the input mirror absorption. As M$_1$ and M$_2$ were coated in the same coating run, identically cleaned and then glued to a closed spacer in a clean room environment their absorption and reflectivity can assumed to be identical, $\alpha_{\rm M_1}=\alpha_{\rm M_2}$ and R$_1$=R$_2$. In s-pol the reflectivity of M$_1$ was found to be $1-R_1=\unit[(529 \pm 20)]{ppm}$. The effective reflectivity of the output mirror was $1-\tilde{R}_{2}=\unit[(2902 \pm 103)]{ppm}$ where $\tilde{R}_{2}$ includes the transmission of M$_2$ as well as the absorption of all three mirrors. Considering known transmission and absorptions, for M$_3$ an upper limit for the absorption of $R_1-\tilde{R}_{2}- 2 \times \alpha_{\rm M_3,s-pol} = \unit[(303 \pm 207)]{ppm}$ remains. Equivalently, for p-polarisation where the reflectivities were $1-R_1=\unit[(4717 \pm 212)]{ppm}$ and $1-\tilde{R}_{2}=\unit[(7882 \pm 251)]{ppm}$ the upper limit for M$_3$ absorption is $\alpha_{\rm M_3,p-pol}=\unit[(309 + 657 / -309)]{ppm}$. An absorption of \unit[(309 + 303)/2 = 306]{ppm} for a highly reflective aSi/SiO$_2$ coating stack at an AOI close to $0^\circ$ is an even more promising result for aSi/SiO$_2$ coatings. The absorption of the incident laser beam would be reduced to \unit[1.1]{ppm} by the upper SiO$_2$/Ta$_2$O$_5$ layers. All values are summarized in Tab.~\ref{tab:abs}.

\section{Discussion and conclusion}

In our model we replaced 11 bilayers of SiO$_2$/Ta$_2$O$_5$ by 5 bilayers of aSi/SiO$_2$ resulting in 26 instead of 38 single layers in total. Due to the high refractive index of $n_{\rm Si}=3.5$ of the aSi layers, using the multi-material stack can achieve same high reflectivity of $T \approx \unit[0.5]{ppm}$ while the total thickness is reduced from $\unit[8.684]{\mu m}$ to $\unit[5.701]{\mu m}$ by \unit[34]{\%}.

Due to reducing the number of bilayers and replacing a part of the Ta$_2$O$_5$ layers by low loss aSi, the loss of the multi-material stack at low temperatures reduces by about \unit[20]{\%} at \unit[120]{K} and \unit[15]{\%} at \unit[20]{K} and \unit[10]{K}. The Brownian thermal noise reduces by about \unit[15]{\%} at RT and \unit[25]{\%} low temperatures.

In our model for a three material coating stack \unit[100]{\%} of the laser power affects the top stack which is made of SiO$_2$/Ta$_2$O$_5$. So we can expect about $\unit[(1.7 \pm 0.3)]{ppm}$ absorption in this part of such a coating. (The largest amount of absorption occurs in the first 2-3 bilayers of a coating stack. So the smaller number of layers compared to the experiment will not reduce the absorption significantly.)
Only \unit[0.35]{\%} of the laser power is transmitted into the aSi/SiO$_2$ part of the stack. This reduces the absorption of the input laser power to 0.0035 x \unit[(1035 +/- 42)]{ppm} = \unit[(3.6 +/- 0.1)]{ppm}. Therefore we conclude that a total optical absorption of as low as $\unit[(5.3 \pm 0.4)]{ppm}$ can be achieved for such a three material based coating stack which is in the order of the coating absorption requirement of about \unit[1]{ppm} of future GWDs.

For an AOI close to $0^\circ$ from the cavity round-trip loss a lower absorption of 0.0035 x \unit[306]{ppm} = \unit[1.1]{ppm} was concluded for aSi/SiO$_2$ coatings which is promising for further absorption reduction.

The aSi in these coating stacks was not optimized to be low absorbing. Work on improving the optical absorption of such aSi layers provides the perspective to replace more Ta$_2$O$_5$ layers by aSi to further reduce Brownian thermal noise. In our design so far the SiO$_2$/Ta$_2$O$_5$ also was not explicitly optimized to have low absorption. Absorptions of less than \unit[0.7]{ppm} at \unit[1064]{nm}~\cite{Beauville2004} suggest the possibility of further reducing the SiO$_2$/Ta$_2$O$_5$ absorption also at \unit[1550]{nm}. 

\section{Acknowledgements}

We acknowledge support from the SFB/Transregio 7, the International Max Planck Research School (IMPRS) on Gravitational Wave Astronomy, and from QUEST, the centre for Quantum Engineering and Space-Time Research.

We are grateful for additional financial support from STFC and the University of Glasgow. IWM is supported by a Royal Society Research Fellowship. SR holds a Royal Society Wolfson Research Merit award. We are grateful to the International Max Planck Partnership for Measurement and Observation at the Quantum Limit for support, and we thank our colleagues in the LSC and VIRGO collaborations and within SUPA for their interest in this work.

This paper has LIGO Document number LIGO-P1400227.

\end{document}